\documentclass[showpacs,pre]{revtex4}
\usepackage{amssymb}

\usepackage{amsmath}
\usepackage{graphicx}
\usepackage{dcolumn}
\usepackage{bm}

\begin{document}

\author{A. Bret}
\email{antoineclaude.bret@uclm.es}

\date{\today }

\title{Fast growing instabilities for non-parallel flows}

 \affiliation{ETSI Industriales, Universidad de Castilla-La
Mancha, 13071 Ciudad Real, Spain}


\begin{abstract}
Unstable modes growing when two plasma shells cross over a background plasma at arbitrary angle $\theta$, are investigated using a non-relativistic three cold fluids model. Parallel flows with $\theta=0$ are slightly more unstable than anti-parallel ones with $\theta=\pi$. The case $\theta=\pi/2$ is as unstable as the $\theta=0$ one, but the fastest growing modes are oblique. While the most unstable wave vector vary with orientation, its growth rate slightly evolves and there is no such thing as a stable configuration. A number of exact results can be derived, especially for the $\theta=\pi/2$ case.
\end{abstract}

\maketitle

\section{Introduction}
Beam plasma instabilities are ubiquitous in physics and have been investigated for many decades \cite{BhomGross}. The topic currently undergoes a renewed interest through the Fast Ignition Scenario for Inertial Fusion \cite{Tabak} or some scenarios of Gamma Ray Bursts production in astrophysics \cite{Medvedev99}. Indeed, Astrophysics offers a very wide range of unstable systems which can be magnetized, relativistic, homogenous or not. Counter-streams  instabilities are usually studied assuming parallel streams. But why should real systems systematically fit this scheme? Admittedly, when an electron beam enters a plasma, a return current is prompted in opposite direction to neutralize it \cite{Hammer}. But two already current and charge neutralized plasma shells could perfectly collide over a background plasma at an arbitrary angle. It seems the problem of non-parallel streams has not be addressed so far, and the goal of this letter is to show that non negligible growth rates can arise from non-parallel streams interactions as well.

\begin{figure}[tbp]
\begin{center}
\includegraphics[width=0.5\textwidth]{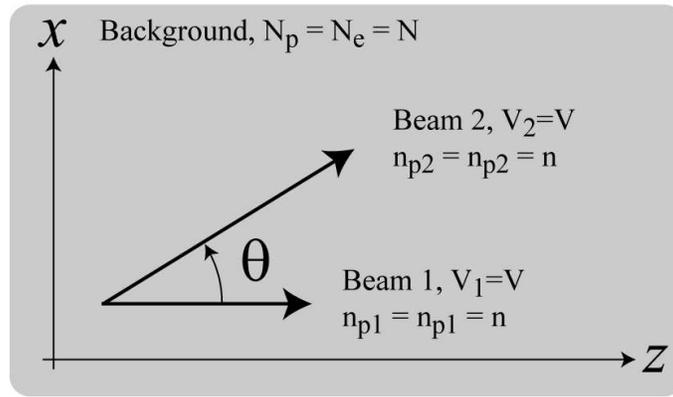}
\end{center}
\caption{Scenario considered.} \label{fig:1}
\end{figure}

Let us consider two non-relativistic beams, both charge and current neutralized. For simplicity, we consider here two electron-proton beams where both species have equal densities so that charge and current neutrality are guarantied regardless of the beams relative motion. The problem of the beams respective orientation is interesting only if there is a background plasma. Otherwise, one just needs to consider the reference frame of one of the beams to cancel any orientation parameter. We eventually come up with the setting pictured of Figure \ref{fig:1}. Calculations are conducted in the reference frame where the background plasma of electronic and protonic density $N$ is at rest. In order to focus on the parameter $\theta$, the two beams have equal protonic and electronic densities $n$, and both flow at the same velocity $V$ along their respective direction. We now study the stability of harmonic perturbations $\propto \exp(i\mathbf{k}\cdot \mathbf{r}-i\omega t)$ where $i^2=-1$ and $\mathbf{k}=(k_x,0,k_z)$. Note that the all $k$ unstable spectrum is evaluated in order to be able to spot the fastest growing modes for any given configuration. Finally, we  neglect proton velocities perturbations in view of their much larger inertia.

\section{Three fluids model}
As a first approximation, we implement a cold three fluids model for the two beams and the plasma. Linearizing the conservation and Euler equations for the three species, the dispersion equation is found evaluating the dispersion tensor in a standard way in terms of the dimensionless variables,
\begin{equation}\label{eq:variables}
   \alpha=\frac{n}{N},~~\mathbf{Z}=\frac{\mathbf{k}V}{\omega_p},~~\beta=\frac{V}{c},~~x=\frac{\omega}{\omega_p},
\end{equation}
where $\omega_p^2=4\pi N e^2/m_e$ is the background plasma frequency, $e$ the electron charge and $m$ the electron mass. Due to the arbitrary orientation of the wave vector and of the two beams, the full dispersion tensor is too large to be reported here. Still, the dispersion equation  remains polynomial and can easily be solved numerically. For $Z_x=0$, the $zz$ component of the tensor give the growth rate of a two-stream like instability, as least for $\theta=0$ and $\pi$. The corresponding dispersion equation reads,
\begin{equation}\label{eq:disperTS}
    1-\frac{1}{x^2}-\frac{\alpha}{(x-Z_z)^2}-\frac{\alpha}{(x-Z_z\cos\theta)^2}=0.
\end{equation}

\begin{figure}[t]
\begin{center}
 \includegraphics[width=0.9\textwidth]{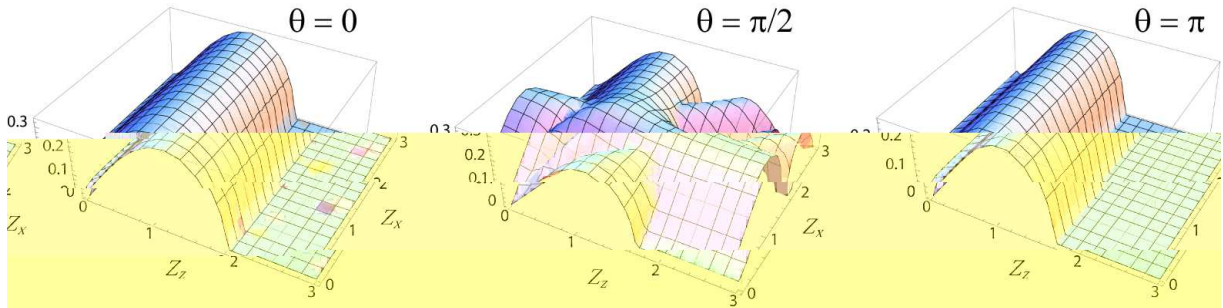}
\end{center}
\caption{(Color Online) Growth rate map in terms of the reduced wave vector $\mathbf{Z}$ for $\alpha=0.1$, $\beta= 0.1$ and $\theta=0$, $\pi/2$ and $\pi$.}\label{fig:2}
\end{figure}

The growth rate map in terms of $\mathbf{Z}$ for $\theta=0$, $\pi/2$ and $\pi$ on Figure \ref{fig:2}. The first and the last cases pertain to well-known systems as $\theta=0$ eventually comes down to one single beam of density $2n$ interacting with the plasma, while $\theta=\pi$ corresponds to two counter-streams crossing over a background plasma. These situations have been well studied, and it is known that within the present non-relativistic regime \cite{BretNIMA}, they are governed by the two-stream instability which dispersion equation is precisely given by Eq. (\ref{eq:disperTS}). The maximum growth rate $\delta_M$ is in the diluted beam regime $\alpha\ll 1$,
\begin{equation}\label{eq:GR-TS}
 \delta_M(\theta=0)\sim \frac{\sqrt{3}}{2}\alpha^{1/3}\omega_p,~~\mathrm{and}~~\delta_M(\theta=\pi)\sim \frac{\sqrt{3}}{2^{4/3}}\alpha^{1/3}\omega_p.
\end{equation}
For $\theta=0$, the two beams act as one, and a resonant unstable mode is feed by the free energy of both at the same time. For $\theta=\pi$, resonant unstable modes can only travel with one single beam, yielding a smaller growth rate.

\section{Maximum growth rate for $\theta=\pi/2$}
Besides these extreme orientations of the two beams, Fig. \ref{fig:2} clearly displays some interesting features for the case $\theta=\pi/2$. One the one hand, the fastest growing mode is here found for $Z_z=Z_x\sim 1$ so that it can stay is phase and exchange energy with both beams at the same time. Such non-trivial orientation of the fast growing mode has been so far related to relativistic effects \cite{fainberg,califano3}. We find here that some unusual system geometry can produce the same effect as the only way for a mode to move in phase with two non-parallel beams is to follow an oblique direction. On the other hand, the maximum growth rate seems very close to the one reached for $\theta=0$, and some finer numerical evaluation shows that it is almost the same. Such equality can be demonstrated setting $\theta=\pi/2$ from the very beginning, and considering the growth rate for wave vectors  fulfilling $Z_z=Z_x$. The dispersion equation for these modes reads,
\begin{equation}\label{eq:pisur2}
    \left[1-\frac{1}{x^2}-\frac{2\alpha}{(x-Z_z)^2}\right]P(x)=0,
\end{equation}
where
\begin{equation}\label{eq:P}
P(x)=(x^2-1)(x-Z_z)^2-2\alpha(Z_z^2+(x-Z_z)^2)-2(x-Z_z)^2Z_z^2/\beta^2.
\end{equation}

Since the first factor in Eq. (\ref{eq:pisur2}) is strictly equal to the dispersion equation (\ref{eq:disperTS}) for $\theta=0$, we here prove that the maximum growth rate for this branch is rigourously the same, and that it is reached for the very same $Z_z$ component. A closer look at $P(x)$ shows that it yields another unstable mode. By setting $Z_z\sim 1$ in the expression of $P(x)$ and developing the result near $x=1$, we find the growth rate of this second oblique unstable mode,
\begin{equation}\label{eq:GR2}
    \delta_{M2}\sim \omega_p\beta\sqrt{\alpha}.
\end{equation}
For $\alpha$ and $\beta$ lower than unity, this secondary growth rate remains smaller than $\delta_M(\theta=0)$.

\begin{figure}[t]
\begin{center}
 \includegraphics[width=0.45\textwidth]{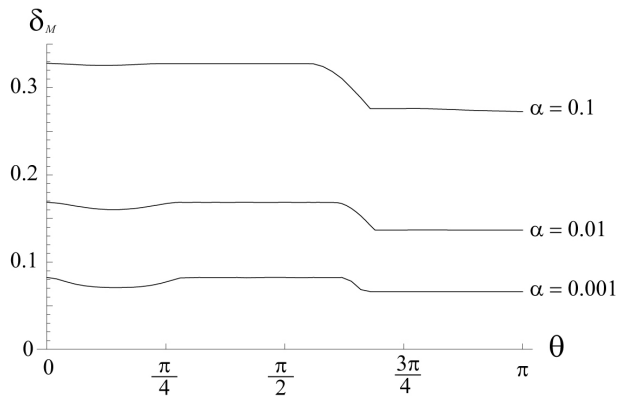}
\end{center}
\caption{Evolution of the maximum growth rate ($\omega_p$ units) for $\beta=0.1$, in terms of the angle $\theta$ and for various beam to plasma density ratios $\alpha$.}\label{fig:3}
\end{figure}

\section{Conclusion}
The overall system is thus found exactly as unstable for $\theta=0$ as it is for $\theta=\pi/2$. Beyond this value of $\theta$, the maximum growth rate must decrease to  its final value $\delta_M(\theta=\pi)$, as given by Eq. (\ref{eq:GR-TS}). Figure \ref{fig:3} shows how $\delta_M$ evolves between $\theta=0$ and $\pi$, for $\beta=0.1$ and various beam to plasma density ratios $\alpha$. As expected, $\delta_M$ remains almost constant between 0 and $\pi/2$ while it falls down to its final value $\delta_M(\pi)$ for $\theta\sim 2$. Because $\delta_M$ is independent of $\beta$ for $\theta=0$, $\pi/2$ and $\pi$, we can expect an overall weak dependance for other angles. This is confirmed through numerical calculation as the three curved displayed on Fig. \ref{fig:3} are indistinguishable from their $\beta=0.01$ counterparts.

Ton conclude, it is important to emphasize that there is no such thing as a stable configuration. On the contrary, the system remains unstable regardless of the beams orientation, and the evolution of the maximum growth rate is limited. The most unstable wave vector is two-stream like for parallel and anti-parallel orientations, but turns oblique in the intermediate case to stay in phase with both beams at the same time. Noteworthy, relativistic and/or kinetic effects should force an oblique regime regardless of the beam orientation so that an evaluation of the whole unstable spectrum becomes mandatory.

\section{Acknowledgements}
This work has been  achieved under projects FIS 2006-05389 of the
Spanish Ministerio de Educaci\'{o}n y Ciencia and PAI-05-045 of
the Consejer\'{i}a de Educaci\'{o}n y Ciencia de la Junta de
Comunidades de Castilla-La Mancha.

\end{document}